\documentclass[prd,aps,a4paper,preprint,preprintnumbers,nofootinbib]{revtex4-1}
\usepackage[a4paper, hdivide={1.91cm,,1.165cm}, vdivide={1.83cm,,3.3cm}]{geometry}

\usepackage{amsmath,amssymb}
\usepackage{graphicx,multirow}
\usepackage{tabularx}
\usepackage[export]{adjustbox}
\usepackage[hyperfootnotes=false]{hyperref}
\usepackage[usenames, dvipsnames]{color}
\usepackage[normalem]{ulem}
\usepackage{float}

\newcommand{\ii}{\ensuremath{\text{i}}}
\newcommand{\dd}{\ensuremath{\text{d}}}

\newcommand{\beq}{\begin{eqnarray}}% can be used as {equation} or  {eqnarray}
\newcommand{\eeq}{\end{eqnarray}}
\newcommand{\bea}{\begin{eqnarray}}% can be used as {equation} or  {eqnarray}
\newcommand{\eea}{\end{eqnarray}}

\newcommand{\del}{\partial}

\begin{document}
%%%%%%%%%%%%%%%%%%%%%%%%%%%%%%%%%

\preprint{UCI-TR-2024-08}

\title{Instabilities of Gauged Q-Balls}

\author{Arvind~Rajaraman}
\email{arajaram@uci.edu}
\affiliation{Department of Physics and Astronomy, 
University of California, Irvine, CA 92697-4575, USA
}

\begin{abstract}

We present the first analytical calculation that shows  that perturbations with angular dependence can lead to an instability in gauged Q-balls.
We find an explicit  condition on the parameters for the Q-ball to become unstable. 
We  compare our predictions to the numerical calculation in Kinach et al., and show agreement, including a correct prediction of the instability/stability of  the two parameter points analyzed in that paper.

\end{abstract}

%%%%%%%%%%%%%%%%%%%%%%%%%%%%%%%%%
\maketitle
\newpage
%\tableofcontents
%%%%%%%%%%%

\section{Introduction\label{s.intro}}

In certain scalar field theories, one can construct solitonic solutions called Q-balls~\cite{Coleman:1985ki}. For these solutions to exist, the scalar field must have a $U(1)$ symmetry, and in addition, the potential must satisfy certain conditions (we review this in Sec. \ref{sec:gauged}). If the symmetry is ungauged, the resulting Q-balls are called global Q-balls, while if the symmetry is gauged, they are called gauged Q-balls~\cite{Lee:1988ag,Gulamov:2013cra,Gulamov:2015fya,Nugaev:2019vru}.

The stability of these solitons has been discussed in many papers, and the picture is still unclear. For global Q-balls, it has been shown~\cite{Lee:1991ax,Friedberg:1976me,Paccetti:2001uh} that the Q-balls are stable as long as  $ {d\omega\over dQ}<0$ (these quantities are defined below). For gauged Q-balls, the analogous statement is known to not be true~\cite{Panin:2016ooo}.

Recently, a numerical analysis of the stability of gauged Q-balls was carried out~\cite{Kinach:2022jdx}. It was found that gauged Q-balls were unstable for certain parameters, and in particular were unstable to modes which had an angular dependence. 
%indeed, for one particular parameter choice that was studied (denoted P2 in that paper), the instability had $L=4$. 
On the other hand, for other choices of parameters, the corresponding Q-balls were stable. In general,  Q-balls  with $ {d\omega\over dQ}<0$ were stable. 

Our goal here is to provide an analytic approach to studying these instabilities. This will be based on the approach to Q-balls developed in~\cite{Heeck:2020bau,Heeck:2021zvk}. These papers showed that in the large radius limit, spherical Q-balls can be analytically described by considering corrections around the thin wall limit (this will be reviewed below in  Sec.~\ref{s.Spherical}). By considering a similar analysis for slightly nonspherical gauged Q-balls, we find  a possible mode of instability for these Q-balls. This analysis is in section~\ref{s.Perturbed}.

We then show in section~\ref{s.Comparison} that the instability that we predict agrees in many details in common with the numerical analysis of~\cite{Kinach:2022jdx}. In particular, we indeed find that for one parameter choice in~\cite{Kinach:2022jdx}, the $L=4$ mode is unstable, in agreement with the numerical results. Similarly, we find that for another parameter choice, the instability is absent, in agreement with the numerical analysis. 
%We also find that Q-balls  with $ {d\omega\over dQ}<0$ are stable. 

We close with comments and 
open questions.

%%%%%%%%%%%%%%%%%%%%%%%%%
%%%%%%%%%%%%%%%%%%%%%%%%%
\section{Q-Balls: Overview and Notation}
\label{sec:gauged}
 
We review here the basic properties of Q-balls; we will follow the notation of~\cite{Heeck:2020bau,Heeck:2021zvk}.
 
The Lagrangian density for a charged scalar field is
\begin{equation}
\mathcal{L}=\left|D_\mu\phi \right|^2-U(|\phi|)-\frac14 F_{\mu\nu}F^{\mu\nu},
\end{equation}
where $D_\mu=\partial_\mu-\ii e A_\mu$ is the gauge covariant derivative and $F_{\mu\nu}=\partial_\mu A_\nu
-\partial_\nu A_\mu$ is the field-strength tensor. 
Q-balls exist when the 
function $U(|\phi|)/|\phi|^2$ has a minimum at $0<\phi_0<\infty$ such that 
\begin{equation}
0\leq\sqrt{\frac{2U(\phi_0)}{\phi_0^2}}\equiv \omega_0< m_\phi\,,\label{e.Omega0}
\end{equation}
where $m_\phi$ is the mass of the complex scalar. 
We make the static charge ansatz~\cite{Lee:1988ag} 
\begin{equation}
\phi (t,\vec{x})=\frac{\phi_0}{\sqrt{2}} f(r) e^{\ii\,\omega t}\,,\qquad A_0(t,\vec{x})=A_0(r)\,,\qquad A_i(t,\vec{x})=0\,,\label{e.fdef2}
\end{equation}
The scalar frequency $\omega$ is  restricted to the region $\omega_0<\omega \leq m_\phi$.

It is convenient to define the dimensionless quantities 
\bea
\rho\equiv r\sqrt{m_\phi^2-\omega_0^2},\quad 
A(\rho)\equiv \frac{A_0 (\rho)}{\phi_0}\,,\quad
 \alpha\equiv e\Phi_0\,,
 \quad 
%&& 
%\Omega_G\equiv\frac{\omega_G}{\sqrt{m_\phi^2-\omega_0^2}}\,, 
\{\Omega_0, \Omega,\Phi_0\} \equiv\frac{\{\omega_0, \omega, \phi_0\} }{\sqrt{m_\phi^2-\omega_0^2}}
\eea
The  equations of motion are 
\bea
\nabla^2 f &&= \frac{1}{\Phi_0^2(m_\phi^2-\omega_0^2)^2}\frac{\dd U}
{\dd f}-\left(\Omega-\alpha A\right)^2f\label{eq.feq}\\
\nabla^2 A  &&= \alpha f^2(A\alpha-\Omega)\label{gaugeeqn}
\eea
The charge is 
\bea
Q=\Phi_0^2\int d^3xf^2\left(\Omega-\alpha A \right) \eea while
the energy is
\bea
E/\sqrt{m_\phi^2-\omega_0^2} =\Phi_0^2\int d^3x\left\{ \frac12f^{\prime 2}+\frac{1}{2}A^{\prime2}
+\frac{1}{2}f^2\left(\Omega-\alpha A \right)^2+\frac{U(f)}{\Phi_0^2(m_\phi^2-\omega_0^2)^2}\right\}
\eea

For concreteness we restrict most of our discussion to the sextic scalar potential 
studied in Ref.~\cite{Heeck:2020bau}. This can be parametrized as
\begin{align}
U(f) = \phi_0^2 \left( \frac{m_\phi^2 -\omega_0^2}{2}\, f^2 (1-f^2)^2 +\frac{\omega_0^2}{2}\, f^2 \right) .
\label{eq:sextic_potential}
\end{align}

\section{ Spherical gauged Q-balls\label{s.Spherical}}

We begin by reviewing the construction of spherical Q-balls.
This section follows the analysis of~\cite{Heeck:2021zvk}.

The spherical Q-ball is approximately a sphere of radius $R$ with a constant internal density. More precisely, in the interior, the scalar field is approximately at the false vacuum solution $f=f_+$ (where 
$f_+$  is the nonzero solution of the equation \ref{eq.feq}) and
at the Q-ball surface, the field rapidly transitions to the true vacuum 
$f=0$.
The scalar profile is thus to a first approximation\bea
f^{(0)}=
\left\{\begin{array}{c}
f_+\qquad \rho<R
\\
0 \qquad \rho>R
\end{array}
\right.
\eea

The gauge field  is a nondynamical field, and is determined from the scalar charge distribution through the Maxwell
equation \ref{gaugeeqn}. For the scalar profile above, 
the solution for the gauge field  is
\bea
\alpha A_{}^{(0)}=
\left\{
\begin{array}{cc}\Omega+Q_{in}{\sinh(\alpha f_+ \rho)\over \rho}&~~~~~~~~~~ \rho<R
\\
{Q_{out}\over \rho}  & ~~~~~~~~~~~~\rho> R
\end{array}
\right.\label{eq.Asol1}
\eea

Note that  the typical scale of variation
of $A$ is inversely proportional to $\alpha$. 

Near $\rho\sim R$, the scalar field is varying very quickly over a short distance of order 1, which we call the transition region. However, since the gauge field varies much more slowly than the scalar, we can take the gauge field to be approximately smooth over the transition region.
 Continuity of the gauge field and its derivative at $\rho=R$ then implies
\bea
Q_{in}=-{\Omega\over {\alpha f_+ \cosh(\alpha f_+ R)}}
\qquad \qquad 
Q_{out}=R\Omega\left(1-{1\over {\alpha f_+ R }}{\tanh(\alpha f_+ R)}\right)\label{eq.Asol3}
\eea

We note that the value of the gauge field at $\rho=R$ is
\bea
\alpha A_{R}^{(0)}=
\Omega\left(1-{1\over \alpha f_+ R}{\tanh(\alpha f_+ R) }\right)
\eea

We can now require consistency between the obtained gauge field and the initial ansatz for the scalar profile. To do this, we solve the equation for the scalar field in the gauge field background (\ref{eq.Asol1}). This equation is now
\begin{align}
{d^2 f^{(0)}\over d \rho^2}+
%(f^{(0)})'' + 
\frac{2}{\rho}{d f^{(0)}\over d \rho} %(f^{(0)})' 
=  f^{(0)}\left(  1-4(f^{(0)})^2+3(f^{(0)})^4
+{\Omega_0^2} 
-\left(\Omega-\alpha A^{(0)}\right)^2\right)\label{eq.transitioneqA}
\end{align}

This is particularly important to solve in the  transition region, where the scalar field varies quickly from the false vacuum to the true vacuum. This requires a very precise relation between the potential parameters and the radius of the Q-ball, as discussed in~\cite{Heeck:2021zvk} for the case of global Q-balls, and which we now review. 

In the gauged Q-balls case that we are discussing here, the scalar field in the transition region would satisfy  
the  equation
\begin{align}
{d^2 f^{(0)}\over d \rho^2} + \frac{2}{R} {d f^{(0)}\over d \rho} =  f^{(0)}\left(  1-4(f^{(0)})^2+3(f^{(0)})^4
+{\Omega_0^2} 
-(\Omega_{R}^{(0)})^2\right)\label{eq.transitioneq}
\end{align}
where we have replaced the explicit factor of $\rho$ by $R$, and set the gauge field to be its value at $\rho=R$. This has allowed us to replace $\left(\Omega-\alpha A^{(0)}\right)$ by the constant
\bea
\Omega_{R}^{(0)}=\left(\Omega-\alpha A^{(0)}_R\right)
={\Omega ~{\tanh(\alpha f_+ R)}\over \alpha f_+ R} \label{Omegarel}
\eea
 %$\Omega_{R}^{(0)}$ is  effectively constant over the transition region.
 
This maps the problem to the form of a global Q-ball equation, and we can use the methods in ~\cite{Heeck:2021zvk} to solve it.

% which is related to $1/R$.
We can treat the term $\frac{2}{R} {d f^{(0)}\over d \rho} $ as a small frictional  term. In its absence, there is an exactly conserved energy, and this allows the equation to be solved exactly; consistency of energy conservation implies that this is a solution in the limiting case  when the 
false vacuum and the true vacuum have the same energy.  In this limit, the solution  of the equation (\ref{eq.transitioneq})  is
\begin{align}
f^{(0)}(\rho)=\frac{f_+}{\sqrt{1+2e^{2(\rho-R)}}} \,.
\end{align}

 If the 
false vacuum and the true vacuum have slightly different energies, the frictional term must be included if the solution is to exist, and furthermore $R$ must have a precise value. Specifically, the 
change in the energy between the two vacua should be balanced by the effect of the  frictional term.
%Wha Since the frequency and the radius are related for a global Q-ball, this equation can only be solved if $\Omega_R^{(0)}=\Omega_{e=0}$, 
%where $\Omega_{e=0}$ is the value of $\Omega$ which would produce a global Q-ball of radius  $R$.  
The details of the derivation of  the resulting consistency condition are in~\cite{Heeck:2021zvk}
and show that the radius must satisfy
\bea
((\Omega_R^{(0)})^2-\Omega_0^2)R=1+{\cal O}( {1\over R })\label{eq.Rrel}
\eea

This can be shown~\cite{Heeck:2021zvk} to be an accurate prediction for the   Q-ball radius.

\section{Perturbed Q-balls\label{s.Perturbed}}

We now look for potential instabilities of these Q-balls.

We do these by considering deformations of the spherical Q-ball. The deformation of the Q-ball will in general not lead to a solution of the equations of motion.
The deformations may grow, indicating that the Q-ball is unstable,
or they may decay away. However, if there is a transition between 
stable Q-balls and unstable Q-balls, then exactly at threshold, the deformation
will neither grow nor decay, and therefore the deformed Q-ball will be a solution of the equations of motion.

Now the spherical Q-ball has an approximately constant density in the interior, and zero density 
outside. We will consider perturbations of the Q-ball where the density  is assumed to 
 be approximately the same constant in the interior, but where the surface deforms in some way.
It is likely that other deformations (e.g. where the internal density is nonconstant) are possible; we leave the search for these to future work. 
The assumptions we make will hence lead to a conservative 
estimate for the region of stability (i.e. our analysis will not find the most general instability, 
but any Q-balls that we find to be unstable will indeed be unstable).

It is fairly immediate that under these assumptions  on the perturbations, a spherical perturbation would change the charge. We must therefore consider perturbations with an angular dependence, such that the  total charge does not change to leading order.

At this point, then, we are led to consider a perturbation where the Q-ball surface is deformed to be 
at 
\bea 
\rho=R(\theta)\equiv R(1+\epsilon P_L(\cos(\theta))).\label{deformedradius}
\eea
 $P_L(\cos(\theta))$ is a Legendre polynomial; we shall denote this as $P_L$, suppressing the $\theta$ dependence. $\epsilon$ parametrizes the size of the perturbation, and can be taken to be parametrically
 small.

We now follow the same approach as the previous section to construct these Q-balls. 

First, we will make an initial ansatz that the scalar field is at the false vacuum in the interior 
%(now deformed as (\ref{deformedradius})) 
and zero outside i.e. 
\bea
f=
\left\{\begin{array}{c}
f_+\qquad \rho<R(1+\epsilon P_L%(\cos(\theta))
)
\\
0 \qquad \rho>R(1+\epsilon P_L%(\cos(\theta))
)
\end{array}
\right.
\label{fsol}
\eea
%assume that the 

%\subsection{The gauge field}

%The first step is to calculate the profile of the gauge field, and in particular the value of the gauge field in the transition region.

%Following the previous section, we  use the thin wall approximation to find the gauge field.
% In the interior we will again have $f=f_+$ while in the exterior $f=0$. 
% 
The equation of motion \ref{gaugeeqn} for the gauge field  can then be solved to find
%in the interior and exterior are therefore unchanged. 
%
% At the linearized level, the gauge field will obtain harmonics proportional to $P_L$. The solution will
%be therefore of the form
\bea
\alpha A
=
\left\{
\begin{array}{c}
\Omega+Q_{in}{\sinh(\alpha f_+ \rho)\over \rho}+
\epsilon P_{in}F_L(\alpha f_+\rho)
P_L\qquad \rho< R(1+\epsilon P_L)
%R(\theta)
\\
~~~~~~~~~Q_{out}{1\over \rho}+
\epsilon P_{out}{1\over \rho^{L+1}}P_L
\qquad~~~~~~~~ \rho> R(1+\epsilon P_L)
%R(\theta)
\end{array}
\right.
\label{eq.Asol2}
\eea
where $F_L(\alpha f_+\rho)=j_L(i\alpha f_+\rho)$, and $j_L$ is a spherical Bessel function. We list the first few $F_L$ in an appendix.

We again require the gauge field and its derivative to be continuous  at the boundary. % $\rho=R(\theta)$. 
This yields 
\bea
Q_{in}=-{\Omega\over {\alpha f_+ \cosh(\alpha f_+ R)}}
\qquad \qquad 
Q_{out}=R\Omega\left(1-{1\over {\alpha f_+ R }}{\tanh(\alpha f_+ R)}\right)
\\
P_{in}={\Omega \tanh(\alpha f R)
\over \alpha fF'_L(\alpha f\rho)
+  F_L(\alpha f R){(L+1)\over \alpha fR}}
\qquad \qquad 
 P_{out}=
 {\Omega R^{L+1} \tanh(\alpha f R)F_L(\alpha f R)
\over F'_L(\alpha fR)
+  F_L(\alpha f R){(L+1)\over \alpha fR}}
\eea
%and $Q_{in}, Q_{out}$ continue to satisfy Eqn. \ref{eq.Asol3}.

At the boundary, the value of the gauge field is now
\bea
\alpha A_R=
%(\theta)=%\alpha A^{(0)}_R
\Omega\left(1-{1\over \alpha f_+ R}{\tanh(\alpha f_+ R) }\right)+\epsilon P_L \Omega\left({  F_L(\alpha fR)\tanh(\alpha f R)
\over F'_L(\alpha f\rho)
+  F_L(\alpha f R){(L+1)\over \alpha fR}}
- 1
+{\tanh(\alpha f R)\over \alpha fR}\right)
\label{eq.ardef}
\eea
%where
%\bea
%a_R=
%{ \Omega F_L(\alpha fR)\tanh(\alpha f R)
%\over F'_L(\alpha f\rho)
%+  F_L(\alpha f R){(L+1)\over \alpha fR}}
%-\Omega \left(1
%-{\tanh(\alpha f R)\over \alpha fR}\right)\label{eq.gammaeq1}
%\eea

%\subsection{The scalar equation}

We can now require again consistency between the obtained gauge field and the initial ansatz for the scalar profile. To do this, we solve the equation for the scalar field in the gauge field background (\ref{eq.Asol2}).
We focus on the  transition region, where the scalar field varies quickly from the false vacuum to the true vacuum.
Once again we note that the gauge field is slowly varying as a function of $\rho$, and can be taken to have  a  value $A_R$ over the transition region. 

The differential equation for the scalar  is
\begin{align}
%-\ddot{f}+
{\del^2 f\over \del \rho^2} + \frac{2}{\rho} {\del f\over \del \rho}+{1\over \rho^2\sin\theta}\del_\theta(\sin\theta \del_\theta f) =  f\left(  1-4f^2+3f^4
+{\Omega_0^2}-\Omega_{R}^2\right)\label{eq.transition7}
\end{align}
where
$\Omega_{R}=\Omega-\alpha A_R$. Since $f$ is now dependent on $\theta$, we have kept the angular derivatives in the Laplacian.
%In the transition region, we  have
%\begin{align}
%-\ddot{f}+
%{\del^2 f\over \del \rho^2} + \frac{2}{R(\theta)} {\del f\over \del \rho}+{1\over \rho^2\sin\theta}\del_\theta(\sin\theta \del_\theta f) =  f\left(  1-4f^2+3f^4
%+{\Omega_0^2}-\Omega_{R}^2\right)\label{eq.transition2}
%\end{align}

This is now a differential equation in two variables, which is obviously much more complex. However, 
we note that all the radial and angular dependence in the ansatz (\ref{fsol}) is at the surface of the Q-ball, and hence these dependences are related. To make this precise, note that the  ansatz (\ref{fsol}) only depends on $\rho-R\epsilon P_L$. We therefore have
%In the ansatz (\ref{eq.fansatz}), the time dependence and the angular dependence are linked to the radial dependence. Indeed, we can write \bea
%-\ddot{f_T}={f_T}' R\gamma{\epsilon}P_L\eea
%and similarly
\bea
{1\over \rho^2\sin\theta}\del_\theta(\sin\theta \del_\theta f)=
{df\over d\rho}{1\over R}L(L+1) \epsilon P_L
\eea
where we have only kept the terms linear in $\epsilon$.
%Note that this is small, and so to leading order we can treat the scalar field as only depending on $\rho$.
The angular derivative terms therefore yield terms proportional to ${df\over d\rho}$. 

In the transition region, we can also set $\rho= R(\theta)=R(1+\epsilon P_L(\cos(\theta)))$. The scalar equation  then becomes
\begin{align}
%-\ddot{f}+
{\del^2 f\over \del \rho^2} + \left(\frac{2}{R(\theta)}+{1\over R}L(L+1) \epsilon P_L\right) {\del f\over \del \rho}
%+{1\over \rho^2\sin\theta}\del_\theta(\sin\theta \del_\theta f) 
=  f\left(  1-4f^2+3f^4
+{\Omega_0^2}-\Omega_{R}^2\right)\label{eq.transition5}
\end{align}

The partial differential equation now only has derivatives with respect to $\rho$. This can then be solved separately for each $\theta$. For example, one can fix $\theta$, and then find the profile by starting at $f=f_+$ at $\rho=0$ and stepping outward to large $\rho$. We therefore get a profile for each value of $\theta$.

%The partial differential equation has become an equation in one variable $\rho$. This is not quite the case, because there is still a dependence on $\theta$ in the coefficient of ${\del f\over \del \rho}$, as well as in the potential. However, since we are considering a small angular perturbation on a large spherical Q-ball, the angular dependence is much slower than the radial dependence, especially in the transition region. We are therefore  justified in solving the differential equation while treating the coefficient of ${\del f\over \del \rho}$, as well as  the potential, as purely functions of $\rho$. 

%Substituting these into equation (\ref{eq.transition2}), we find that in the transition region, the equation is
%\begin{align}
%f'' + \frac{2}{R_{eff}} f'
%=  f\left(  1-4f^2+3f^4
%+{\Omega_0^2}
%-\Omega_{R}^2\right)\label{eq.transition3}
%\end{align}
%with 
%\bea
%\frac{2}{R_{eff}}=\frac{2}{R(\theta)}+R\gamma \epsilon P_L+\epsilon {1\over R}L(L+1) P_L
%\eea

For each value of $\theta$, the equation is again of the form of  a global Q-ball equation. 
We can write the equation (\ref{eq.transition5}) as
\begin{align}
{d^2 f\over d \rho^2}  + \frac{2}{R_{eff}} {d f\over d \rho}
=  f\left(  1-4f^2+3f^4
+{\Omega_0^2}
-\Omega_{R}^2\right)\label{eq.transition6}
\end{align}
where we have defined 
\bea
\frac{2}{R_{eff}}=\frac{2}{R(\theta)}+\epsilon {1\over R}L(L+1) P_L
\eea
This is identical to the equation (\ref{eq.transitioneq})in the previous section
with the replacement $\Omega^{(0)}_{R}\to \Omega_{R},
R\to R_{eff}$. The solution may therefore be read off from (\ref{eq.Rrel}) and states that
the perturbed Q-ball satisfies
%The relation  is therefore modified at large radius to
\bea
\Omega_{R}^2-{\Omega_0^2}  ={1\over R_{eff}}+ {\cal O}({1\over R^2_{eff}})
\eea

This may be simplified by matching terms linear in $\epsilon$ on both sides, and yields the equation for a stable deformed Q-ball as
%Eliminating $R_{eff}$, we find
%\bea
%2(\Omega_{R}^2-{\Omega_0^2})=
%2((\Omega^{(0)}_{R})^2-{\Omega_0^2})+R\gamma \epsilon P_L+\epsilon {1\over R}(L(L+1)-2) P_L
%\eea
%This implies
%We set $\alpha A_R=\alpha A_R^0+\epsilon a_RP_L$, to obtain
\bea
 { \Omega F_L(\alpha fR)\tanh(\alpha f R)
\over F'_L(\alpha f\rho)
+  F_L(\alpha f R){(L+1)\over \alpha fR}}
-\Omega \left(1
-{\tanh(\alpha f R)\over \alpha fR}\right)=-{(L+2)(L-1)\over 4\Omega^{(0)}_{R} R}
%R\gamma \label{eq.gammaeq2}
\eea
%where $a_R$ is defined in (\ref{eq.ardef}).
This is also the relation that gives the instability threshold. That is, if there is a transition from a region of stable Q-balls  to a region of unstable Q-balls, the boundary is set by this relation.

What if this relation is not satisfied? Then the deformed Q-ball does not satisfy the equations of motion, and it will either relax back to the spherical Q-ball, or the deformation grows, indicating an instability.  At least close to the threshold, we can track this by giving $\epsilon$ a small time dependence. Following the same derivation, we 
find
\bea
{R\ddot{\epsilon}\over \epsilon} =-4\Omega \Omega^{(0)}_{R}
\left({  F_L(\alpha fR)\tanh(\alpha f R)
\over  F'_L(\alpha f\rho)
+  F_L(\alpha f R){(L+1)\over \alpha fR}}
- \left(1
-{\tanh(\alpha f R)\over \alpha fR}\right)\right)-{(L+2)(L-1)\over R}\label{finresult}
\eea
Therefore, if the right-hand side of the above equation is negative, the Q-ball is stable, but if it is
 positive, there is a instability.

\section{Comparison with Numerics\label{s.Comparison}}

%Equation () gives us a relation for the instability threshold; 
We will now compare this theoretical prediction for the instability threshold of gauged Q-balls to the numerical results of ~\cite{Kinach:2022jdx}.

%\subsection{The point P2}

The authors of~\cite{Kinach:2022jdx} consider Q-ball instabilities in several types of potentials; 
one of the models they consider is a gauged Q-ball in a sextic potential with  
\bea
V=|\phi|^2-{1\over 2} |\phi|^4+{0.2\over 3} |\phi|^6
\eea

For this potential they present three results:

(i) For a particular parameter point  P2 (details below) the Q-ball is unstable. The instability has $L=4$.

(ii)  For a particular parameter point  P1 (details below) the Q-ball is stable.

(iii) For a particular choice of parameters, they find the instability threshold.

We review these results, and compare to our prediction.

(i) The parameter point  P2 has $\omega=0.9958, e=0.02, r=65$. After rescaling to match our conventions, we find that this parameter point corresponds to
\bea
\Omega_0^2=1/15
\qquad
\Omega=1.028
\qquad
R=62.9
\qquad
\alpha=0.057 \qquad
%\alpha R=3.56 \qquad
%\eea
%From (\ref{eq.omegas}), we also have  
%\bea
\Omega^{(0)}_{R}
%=\left(\sqrt{1/15+{1\over 62.9^2}}\right)
=0.2587
\eea

For these parameters, we calculate the right-hand-side of eqn (\ref{finresult})  for each value of $L$. 
We find that the right-hand-side vanishes for $L=1$; this reflects the translational mode of the Q-ball, and must
be ignored.
%We now calculate the value of $R\gamma$ for each $L$. The results are shown in  Table \ref{t.table1}. 
%
%
%\begin{table}[h!]
%\begin{tabular}{|c|c|c|c|c|c|c|}
%\hline
%L&1&2&3&4&5&6\\\hline
%$R\gamma$&0&0.123&0.148&0.102&-0.0001&-0.15
%\\
%\hline
%\end{tabular}
%\caption{$R\gamma$ as a function of $L$ for the point P2}\label{t.table1}
%\end{table}
%
%We first note that $L=1$ appears to be a mode which is marginal i.e. neither stable nor unstable. This mode in fact corresponds to a translation of the Q-ball, and will be henceforth ignored.
The right-hand-side is positive for 
 %We see from the above that 
$L=2,3,4$ 
%all have $\gamma>0$, 
and hence  all these modes are predicted to  lead to an instability. The value 
of the RHS for $L=5$ is negative, but very close to zero, and  this result may be changed in 
a more accurate analysis. For $L>5$, the right-hand-side is negative, and hence these modes
do not lead to an instability . 

This is to be compared with the results of~\cite{Kinach:2022jdx}, which finds that the Q-ball with this parameter choice is unstable, with an instability mode with $L=4$. They also find the instability is near the surface of the Q-ball.  This is all consistent with our results. 
%The only discrepancy is that we predict an $L=3$ mode which is in some sense more unstable; however, the absence of this mode, while somewhat puzzling, may plausibly be attributed to some feature of the numerical analysis.

%\subsection{Location of the Instability Threshold} We now look at the predicted instability threshold for 

(ii) The parameter point P1 discussed in~\cite{Kinach:2022jdx}  has $e=0.17, \omega=0.9976$.
This corresponds in our notation to $\Omega=1.029, \alpha=0.481$. The value of $\Omega_0$ is again $1/\sqrt{15}$.

For such large values of $\alpha$, the analytical prediction of the previous section cannot be trusted. Nevertheless, we can see what the formula predicts.
We find that the lowest instability threshold (i.e. the smallest unstable $R$) is predicted to occur for 
$L=2$ at a value 
%For this value of alpha
%RGammPred2[x]=0 at 
$\alpha R=6.7$. This corresponds to a predicted threshold radius of about 14 (that is, Q-balls with a radius smaller than 14 are predicted to be stable, while larger Q-balls are predicted to be unstable).

 The Q-ball radius at parameter point P1 is not given in~\cite{Kinach:2022jdx}, but we may compare the charges at these two parameter points (the charge for P1 is given as 387.5 while the charge for P2 is given as $1.480\times 10^6$) to estimate the ratio of the raidii for these two parameter points. 
%the same potential was considered in 
%\cite{Loginov:2020xoj}, and  
This shows that the radius at parameter point P1 should be about 4. Our prediction would therefore be 
 that the Q-ball at parameter point P1 is stable. This is 
in agreement with the numerical results.

(iii) 
For $e=0.02$ (corresponding to  $\alpha=0.057$ in our conventions), the authors of \cite{Kinach:2022jdx} also present the value 
of the instability threshold i.e. the first value of $R$ for  which the Q-ball is unstable. 
This is found to be very close
to the radius where ${d\Omega\over dR}$ becomes positive. This is at $\alpha R=0.51$.

Let us see what the equation (\ref{finresult}) predicts. Unfortunately, we do not have the other parameter values of the numerical calculation, and we are forced to make estimations of these.
Specifically, we need to find $\Omega , \Omega^{(0)}_{R}$ at the threshold.
% a function of $R$. 
%To do this, we calculate $R\gamma$ as a function of $R$. For 

We first calculate $\Omega^{(0)}_{R}$ using (\ref{eq.Rrel}) to find
\bea
\Omega^{(0)}_{R}=\left(\sqrt{1/15+{1\over R^2}}\right)
\eea
For $\Omega$, we use the known values at parameter point P2, as well as equation (\ref{Omegarel}) to find $\Omega$ as
\bea
{\Omega\over 1.028}={\Omega_{R}^{(0)}\over 0.2587}
{ \alpha f_+ R_t\over{\tanh(\alpha f_+ R_t)} } {\tanh(3.26)\over 3.26}
\eea
Substituting $\Omega , \Omega^{(0)}_{R}$ into equation (\ref{finresult}) gives us an equation determining the instability threshold  for each $L$. 
We  find that the smallest radius of instability occurs for $L=2$, and for a predicted value of   $\alpha R=1.99$. 
%This corresponds to a radius of about 26.5, and w
We therefore predict that all radii above this value are unstable.

%We compare this to the radius where ${d\Omega\over dR}$ becomes positive. This is at $\alpha R=0.51$, and it can then be checked straightforwardly, that for the radii which are in the unstable regime, we indeed have ${d\Omega\over dR}>0$.

Here we have a significant difference between our predicted instability threshold of $\alpha R=1.99$, and the numerically calculated threshold~\cite{Kinach:2022jdx} of $\alpha R=0.51$. We discuss this further in the next section.

\section{Discussion and Conclusion}

We have constructed an explicit perturbation of a gauged Q-ball, and shown that in some parameter ranges, the equations of motion directly show that these perturbations 
can be unstable. These perturbations have angular dependence, and are localized near the surface of the Q-ball. 
We have found an explicit prediction for the condition on the radius (and other parameters) for the modes to become unstable. 

This is, to our knowledge, the first calculation that shows analytically that perturbations with angular dependence can lead to an instability in gauged Q-balls.

We have compared our predictions to the numerical calculation in~\cite{Kinach:2022jdx}, and shown that we correctly predict the instability/stability of  the two parameter points analyzed in that paper. Furthermore, it was found in the numerical analysis that one of the points had an unstable mode with $L=4$, and our formula is able to reproduce this by showing that the $L=4$ mode is indeed unstable. Another parameter point was found to be stable, and this is reproduced by our formula. We also find that increasing the gauge coupling makes the Q-balls more stable, a surprising result which was also found in the numerics.

Finally, we note that our predicted  boundary between stable and unstable parameters does not agree perfectly with the results of~\cite{Kinach:2022jdx}. 
It appears that any Q-ball found to be unstable by our formula is also found to be unstable in the numerical calculations; however, there appear to be Q-balls which are found numerically to be unstable which are not captured by our analysis. This suggests strongly that there are other  perturbations that we have missed that will extend the region of instability.
On a related note, the region of instability in the numerical results appears to be linked to the condition ${d\omega\over dQ}>0$, at least for small gauge couplings. We do not see such a link, which is again suggestive that a different form of perturbation may be relevant for  Q-balls. 
This analysis is left for future work.

\section{Acknowledgments}
This work was supported in part by NSF Grant No.~PHY-2210283. We thank M.~Smolyakov for a comment on an initial version of this paper.

\section{Appendix A: Radial solutions}
For completeness, we list the first few $F_L$ solutions used in the paper 
\bea
F_0&&={\sinh x\over x}
\\
F_1&&=\sinh x\left({1\over x^2}\right)
-\cosh x\left({1\over x}\right)
\\
F_2&&=-\cosh x\left({3\over x^2}\right)
+\sinh x\left({1\over x}+{3\over x^3}\right)
\\
F_3&&=
-\cosh x\left({1\over x}+{15\over x^3}\right)
+\sinh x\left({6\over x^2}+{15\over x^4}\right)
\\
F_4&&
=-\cosh x\left({10\over x^2}+{105\over x^4}\right)
+\sinh x\left({1\over x}+{45\over x^3}
+{105\over x^5}\right)
\\
F_5&&
=-\cosh x\left({1\over x}+{105\over x^3}
+{945\over x^5}\right)
+\sinh x\left({15\over x^2}+{420\over x^4}+{945\over x^{6}}\right)
\\
F_6&&
=-\cosh x\left({21\over x^2}+{1260\over x^4}
+{10395\over x^6}\right)
+\sinh x\left({1\over x}+{210\over x^3}+{4725\over x^{5}}+{10395\over x^{7}}\right)
\eea

\bibliographystyle{utcaps_mod}
\bibliography{BIB}

\end{document}